\documentclass[12pt,preprint]{aastex}

\shorttitle{Polarized Emission in Sagittarius A*}

\begin{document}


\title{The Nature of Linearly Polarized Millimeter and Sub-millimeter Emission in Sagittarius A*}


\author{Siming Liu,\altaffilmark{1} Lei Qian,\altaffilmark{2} Xue-Bing Wu,\altaffilmark{2}
Christopher L. Fryer,\altaffilmark{1, 3} and Hui Li\altaffilmark{1}
}

\altaffiltext{1}{Los Alamos National Laboratory, Los Alamos, NM 87545; liusm@lanl.gov, hli@lanl.gov}
\altaffiltext{2}{Department of Astronomy, Peking University, Beijing 100871; qianl@vega.bac.pku.edu.cn}
\altaffiltext{3}{Physics Department, The University of Arizona, Tucson, AZ 85721; clfreyer@lanl.gov}


\begin{abstract}

The linearly polarized millimeter and sub-millimeter emission in Sagittarius A* is produced within 10 Schwarzschild radii of 
the supermassive black hole at the Galactic Center and may originate from a hot magnetized accretion disk, where electrons 
are heated efficiently by turbulent plasma waves. In such a scenario, the flux density and polarization are very sensitive to 
the electron heating rate and the inclination angle of disk, respectively, and the major axis of the sub-millimeter intrinsic 
polarization, which is aligned with the rotation axis of the disk, is perpendicular to the major axis of the polarized 
near-infrared emission. In combination with MHD simulations, which study the properties of the magnetic field and viscous 
stresses, the current spectral and polarization measurements give tight constraints on the model parameters. Simultaneous 
observations will be able to test the model.

\end{abstract}





\section{Introduction}

Our understanding of physical processes in Sagittarius A*, the compact radio source associated with a 
supermassive [$M\simeq 3.7 \times 10^{6}M_\odot$] black hole at the Galactic Center (Sch\"{o}del et al. 2002; 
Ghez et al. 2005a), has improved dramatically since the detection of linear polarization of its millimeter and 
sub-millimeter emission (Aitken et al. 2000). It is generally accepted that Sagittarius A* is powered by 
accretion of the black hole in stellar winds (Melia 1992;  Narayan 1998; Rockefeller et al. 2004). The linear 
polarization reveals a synchrotron origin of the emission and sets strict constraints on the magnetic field, gas 
density, and electron temperature (Agol 2000; Melia et al. 2000; Quataert \& Gruzinov 2000). Recent VLBI imaging 
shows that the 3.5 mm emission is produced within $10\ r_S$ of the black hole, where $r_S \simeq 1.1\times 
10^{12} [M/(3.7\times 10^6M_\odot)]$ cm is the Schwarzschild radius for a non-spinning black hole (Shen et al. 
2005). The emission may originate from a hot magnetized accretion disk, and simultaneous spectrum 
and polarization measurements can be used to estimate the model parameters (Bromley et al. 2001).

High spatial and spectral resolution X-ray and near-infrared (NIR) observations routinely detect flares from the 
direction of Sagittarius A* (Baganoff et al. 2001; Genzel et al. 2003). The characteristic variation time scale 
of a few tens of minutes, comparable to the dynamical time at the last stable orbit, indicates that they are 
also produced within a few $r_S$ of the black hole. The spectra, polarization, and variability, especially when 
considering results from simultaneous multi-wavelength observations, suggest that the NIR radiation is emitted 
through synchrotron processes and the X-rays are likely due to the synchrotron self-Comptonization (Eckart et al. 
2006a, 2006b;  Yusef-Zadeh et al. 2006a). The flares can be triggered by rapid releases of magnetic field energy 
near the black hole, which then heats electrons to a few tens of MeV producing the observed emission (Liu et al. 
2004; Bittner et al. 2007). The long wavelength radio emission is less variable. The low quiescent-state X-ray 
flux and high centimeter radio flux density uncover a non-thermal origin of the emission and/or emission from an 
unbounded flow at large radii (Liu \& Melia 2001; Yuan et al. 2002). Correlated flare activities in the X-ray and radio bands 
suggest that the radio emission is associated with an outflow (Zhao et al. 2004; Yusef-Zadeh et al. 2006b).

Recent SMA and BIMA observations confirm the linear polarization of the millimeter and sub-millimeter emission 
(Bower et al. 2005; Marrone et al. 2006a, 2006b). These observations not only show that the flux density and 
polarization are variable on time scales as short as a few hours but also reveal a relatively high mean flux 
density and hard spectrum in the sub-millimeter band as compared with the millimeter spectral bump first 
observed by Falcke et al. (1998). The observed position angle of the polarization vector has been used to infer 
an external Faraday rotation measure of $\sim -50$ rad cm$^{-2}$ (Macquart et al. 2006; Marrone et al. 2007) and a 
position angle of $\sim 170^\circ$ for the intrinsic polarization, which appears to be $90^\circ$ higher than the 
position angle of the polarized NIR emission (Eckart et al. 2006b; Meyer et al. 2006).

In this {\it Letter}, we consider a Keplerian accretion flow in a pseudo-Newtonian potential. Instead of assuming that the 
electron and proton temperatures are the same (Melia et al. 2001), we describe the electron heating by turbulence with a 
single parameter and treat the cooling process more accurately. The ratios of the viscous stress to the 
magnetic and gas pressures can be determined by MHD simulations. These leave three basic model parameters, namely the 
inclination angle of the disk, accretion rate, electron heating rate. \S\ \ref{eqs} gives the basic equations describing the 
disk structure. We then discuss the synchrotron emission from the disk and its polarization and apply the model to 
Sagittarius A* in \S\ \ref{app}. In \S\ \ref{dis}, we summarize the main predictions, discuss the model limitation, and draw 
conclusions.

\section{Basic Equations for the Disk Structure}
\label{eqs}



The basic equations for a Keplerian two-temperature accretion flow in a pseudo-Newtonian potential with a 
magnetic field dominated by the azimuthal component are described by Liu et al. (2007). The scale height of the 
disk is given by
\begin{equation}
H = \left[{r k_B(T_p+T_e)(1+2\beta_p)\over GMm_p}\right]^{1/2}(r-r_S)\,,
\end{equation}
where $G$, $k_{\rm B}$, $m_p$, $M$, $r$, $r_S\equiv 2GM/c^2$, $c$, $T_e$ and $T_p$ denote the gravitational 
constant, Boltzmann constant, proton mass, black hole mass, radius, Schwarzschild radius, speed of light, 
electron and proton temperatures, respectively. $\beta_p=B^2/8\pi nk_B(T_e+T_p)$ is the ratio of the magnetic 
field energy density $B^2/8\pi$ to the gas pressure $nk_B(T_e+T_p)$, where $n$ gives the gas density. 
>From the angular momentum conservation, one obtains the radial velocity
$
v_r = -{\beta_\nu\beta_pk_B(T_p+T_e)(r-r_S)/ [m_p(GMr)^{1/2}}]\,,
$
where $\beta_\nu$ is the ratio of the total 
stress to $B^2/8\pi$ (Melia et al. 2001)
and we have assumed that there is no angular momentum flux through the disk.
So
\begin{equation}
n = {GM\dot{M} m_p^{1/2}\over 4\pi \beta_\nu\beta_p [k_B(T_p+T_e)]^{3/2}(1+2\beta_p)^{1/2}r(r-r_S)^2}\,.
\end{equation}

The energy conservation equation is given by
\begin{equation}
{{\rm d} \over {\rm d} r}\left\{k_{\rm B}[T_e(\alpha+1+2\beta_p)+T_p(2.5+2\beta_p)]+m_p\left[(1-2f){v_{\rm K}^2\over 2}+{v_r^2\over 2}-{GM\over r-r_S}\right]\right\} = -{\Lambda\over v_r n}\,,
\label{energy}
\end{equation}
where $\Lambda$ is the radiative cooling rate and $v_{\rm K}= (GMr)^{1/2}/(r-r_S)$ is the Keplerian velocity. $\alpha = 
x[3K_3(x)+K_1(x)-4 K_2(x)]/4K_2(x)$, where $x=m_ec^2/k_B T_e$, $K_i$ refers to the $i$th order modified Bessel function and 
$m_e$ represents the electron mass. To obtain the disk structure, one also needs to specify the electron heating rate by the 
turbulent magnetic field (Liu et al. 2007; Sharma et al. 2007a). The electron heating time $
\tau_{ac} = {3C_1 H \langle v_e\rangle /c_S^2}\,,
$
where $C_1$ is a dimensionless constant, $\langle v_e \rangle={2c (1+x)/x^2 K_2(x)\exp(x)}$  and $c_S = [k_B(T_i+T_e)(1+2\beta_p)/m_p]^{1/2}$ are the mean speed of electrons and speed of fast mode waves, respectively. Then we have
\begin{equation}
{{\rm d}T_e\over {\rm d}r} = {T_e\over \tau_{ac} v_r}+{T_p-T_e \over \tau_{\rm Coul} v_r} -{\Lambda \over \alpha n k_Bv_r}\,,
\label{energye}
\end{equation}
where 
$\tau_{\rm Coul} = {3\pi m_em_p\langle v_e\rangle ^3/ 256 n e^4\ln \lambda}$
is the electron-proton energy exchange time via Coulomb collisions and $\ln\lambda\simeq15$ (Spitzer 1962). 

The cooling is dominated by synchrotron, inverse Comptonization and bremsstrahlung processes. 
In the optically thin regime, the synchrotron and bremsstrahlung cooling rates are given, respectively, by
\begin{eqnarray}
\Lambda^0_{\rm syn} &=& {4e^4n\over 9 m_e^4 c^5} \langle p^2 \rangle B^2 \simeq 1.06\times 10^{-15}n B^2 {3x^2+12x+12\over 
x^3+x^2}\,,
\label{syn0} \\
\Lambda_{\rm brem} &=& \left({2\pi k_{\rm B} T_e\over 3 m_e}\right)^{1/2} {32\pi e^6\over 3 h m_e c^3} n^2 g_{\rm
B} =1.4\times 10^{-27} T_e^{1/2} n^2 g_{\rm B}\,,
\end{eqnarray}
where $\langle p^2\rangle $,  $g_{\rm B}\simeq 1.2$, $e$, and $h$ are the mean momentum square of the electrons, Gaunt factor, elemental charge unit, and Planck constant, respectively, and all quantities here and in what follows are given in cgs units. Most of the thermal synchrotron emission is emitted at
$\nu_E \simeq 20 \nu_c = {60eB(x+1)^2/ 4\pi m_e c x^2}=8.6\times 10^7 B{(x+1)^2/x^2}
\,  {\rm Hz}$ (Liu et al. 2006). The optical depth at $\nu_E$ can be approximated as
$\tau_E 
\simeq {3(x+1)H \Lambda^0_{\rm syn} c^2 /[8(x+10)\pi k_{\rm B} T_e \nu_E^3]}
\,,$
where 
the numerical factor is chosen such that the expression is accurate in the relativistic regime. When $\tau_E\ge 1$, the synchrotron cooling is suppressed due to self-absorption:
\begin{equation}
\Lambda^{\tau}_{\rm syn} \simeq {8(x+10)\pi k_{\rm B} T_e \nu_E^3\over 3(x+1) c^2 H}\,.
\label{syntau}
\end{equation}
We consider the synchrotron self-Comptonization cooling, then the total cooling rate
\begin{eqnarray}
\Lambda &=& \Lambda_{\rm syn} + \Lambda_{\rm IC} + \Lambda_{\rm brem} 
=\Lambda_{\rm syn}(1-8\pi\Lambda_{\rm syn}H/cB^2)^{-1} +\Lambda_{\rm brem} 
\,. 
\end{eqnarray} 
where $\Lambda_{\rm syn}$ is given by equations (\ref{syn0}) and (\ref{syntau}) for $\tau_E<1$ and $\ge1$, respectively. 

\begin{figure}[bht] 
\begin{center}
\includegraphics[height=8.0cm]{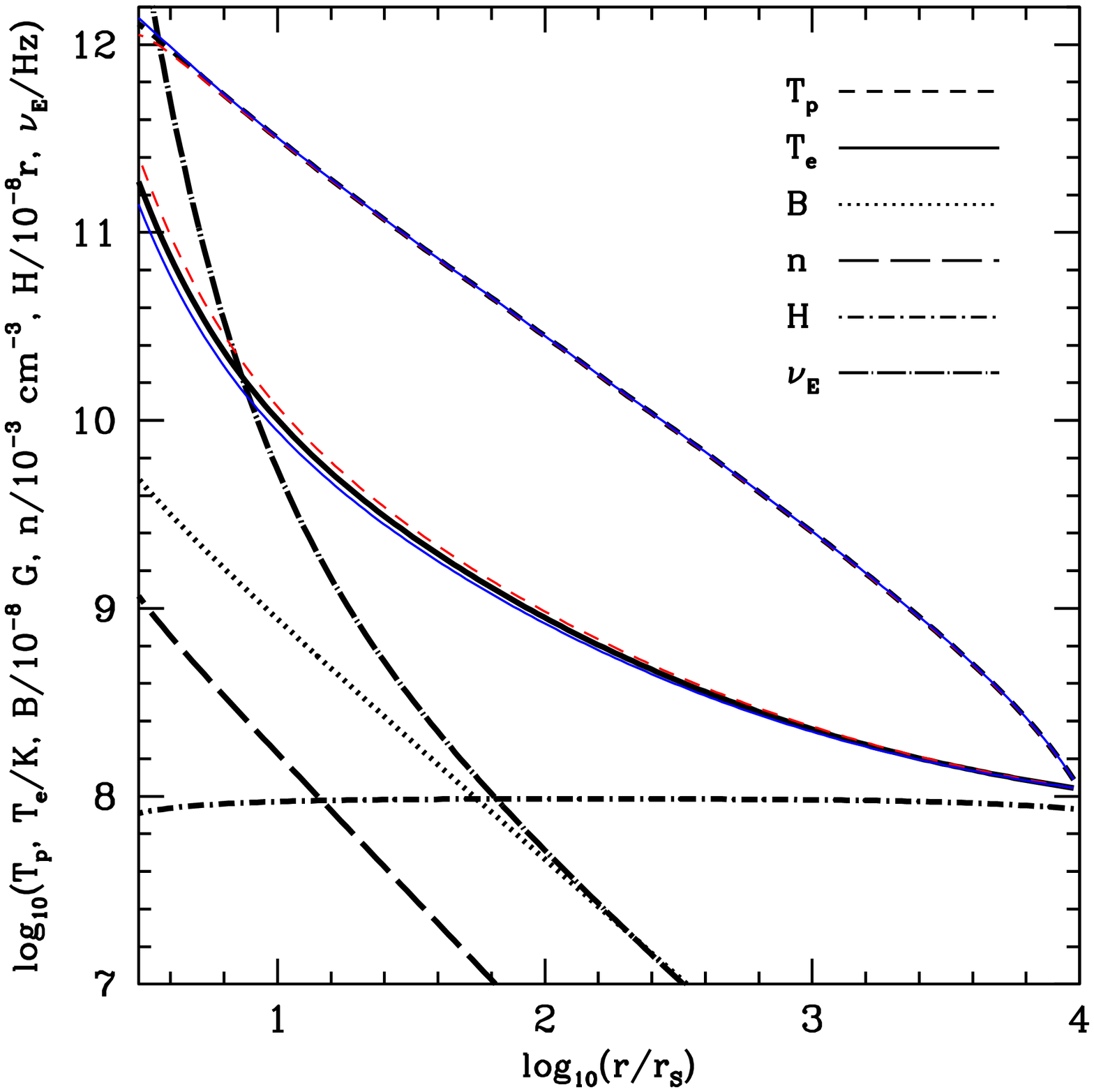}
\hspace{-0.0cm}
\includegraphics[height=8.0cm]{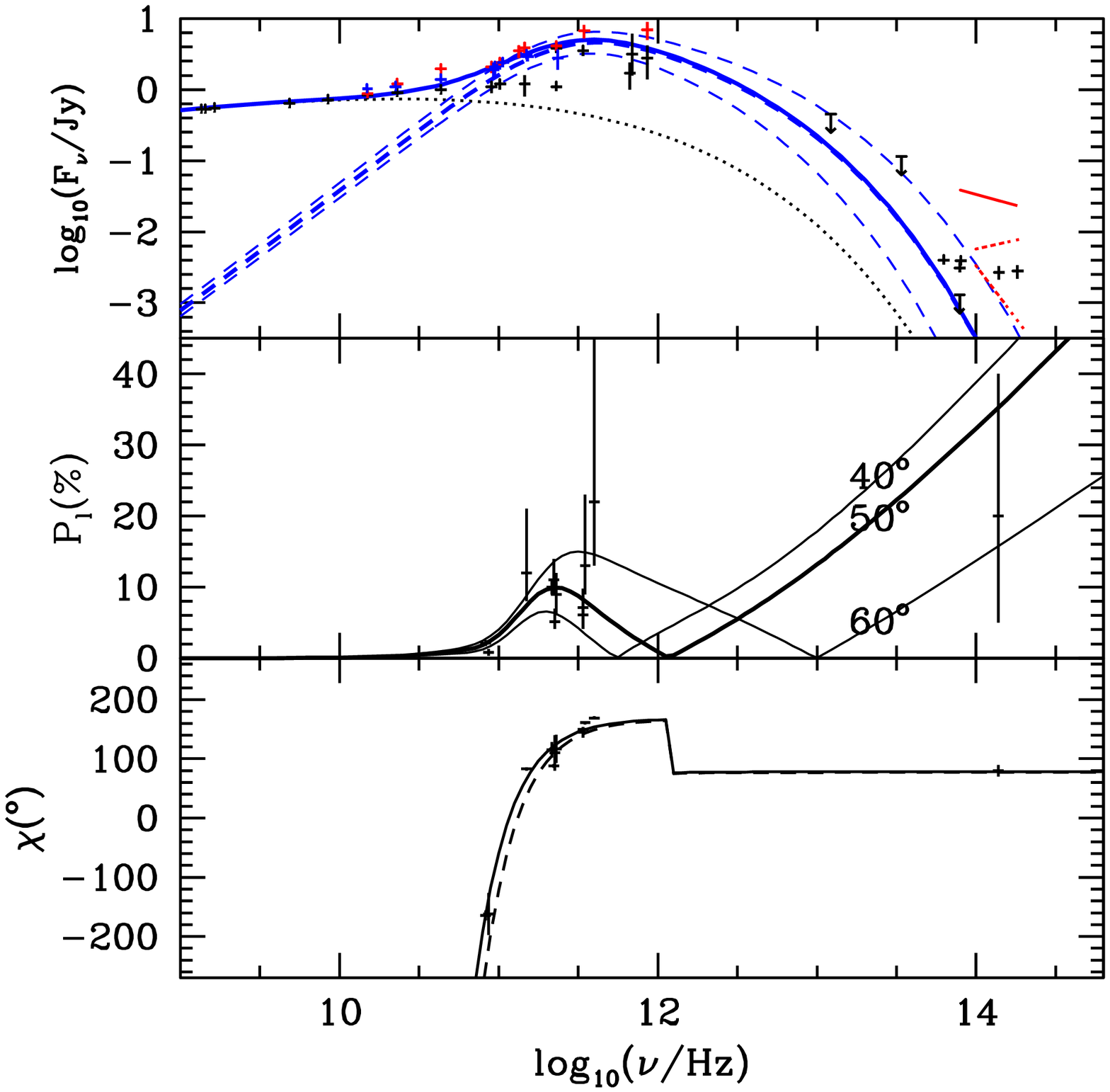}
\end{center}
\caption{
{\it Left:} Profiles of the electron (solid), proton (dashed) temperatures, magnetic field (dotted), gas density (long 
dashed), scale height (dot-dashed), and $\nu_E$ (dot-long-dashed) for the fiducial model with $\beta_\nu = 0.7$, $\beta_p = 
0.4$, $C_1 = 0.38$, $\dot{M}=4.0\times 10^{17}$ g s$^{-1}$, $i = 50^\circ$, and $T_e=T_p=GMm_p/5 k_{\rm B} r$ at the outer 
boundary. The thin dashed and solid lines show the temperature profiles for $C_1=0.36$ and $0.40$, respectively. The other 
parameters don't change, and profiles of other quantities change very little. 
{\it Right:} Model fit to the broadband spectrum and linear polarization. The {\it top} panel shows the fitting to the 
spectrum (Liu et al. 2004). The spectra of NIR flares are from recent observations [Gillessen et al. 2006 (dotted); Hornstein 
2007 (solid)]. The dotted line corresponds to an unpolarized component with $F_\nu = (\nu/10^{9.5}{\rm 
Hz})^{0.3}\exp[-(\nu/10^{10.5}{\rm Hz})^{1/3}]$ Jy. The thick dashed line fits the millimeter to NIR spectrum with the 
fiducial 
model. The thick solid line is the sum of the two. The thin dashed lines are for the two profiles with $C_1=0.36$ (with a 
higher flux) and $0.40$. The {\it middle} panel compares the polarization fractions. The data are from Aitken et al. (2000), 
Macquart et al. (2006), and Marrone et al. (2007). The thick line is for the fiducial model. The thin lines have $i=60^\circ$ 
(with a higher millimeter and sub-millimeter polarization) and $40^\circ$. {\it Bottom}: the observed position angle of the 
linear polarization for the fiducial model. The rotation axis of the disk has a position angle of $168^\circ$. The solid and 
dashed lines assume an external Faraday rotation measure of -44 and -56 rad cm$^{-2}$, respectively (Macquart et al. 2006; 
Marrone et al. 2007). According to the model, the position angle of the polarization flips by $90^\circ$ in the 
sub-millimeter range at the frequency,  where the polarization fraction approaches to 0.  
}
\label{f1.eps}
\end{figure}

For given $\beta_\nu$, $\beta_p$, $\dot{M}$, $C_1$, and $T_e$ and $T_p$ at an outer boundary, one can solve equations 
(\ref{energy}) and (\ref{energye}) numerically to obtain the disk structure. The thick lines in the left panel of Figure 
\ref{f1.eps} give profiles of the fiducial model, where $\beta_\nu=0.7$, $\beta_p=0.4$, $\dot{M}=4.0\times 10^{17}$g 
s$^{-1}$, $C_1=0.38$, and $T_e=T_p=GMm_p/5 k_{\rm B} r$ at the outer boundary $r=10^4 r_S$. For such a low accretion rate, 
cooling is not very efficient and the temperature profiles are determined by the electron heating rate. The thin solid and 
dashed lines give the temperature profiles for $C_1=0.40$ and $0.36$, respectively. The other model parameters remain the 
same. Due to the increase of the electron heating rate, the electron temperature is higher for the latter, which has a 
slightly lower proton temperature for energy conservation. The profiles of other quantities do not change significantly.
We note that the disk structure at small radii is very sensitive to the stress at the inner boundary $3r_S$. Here the stress 
is chosen such that there is no net angular momentum flux through the disk, which is appropriate for strongly magnetized 
disks. 

\section{Radiation Spectrum and Polarization}
\label{app}

The formulae for calculating polarized synchrotron emission are given by Melia et al. (2001), who treat the flow as a slim 
disk. Since the millimeter and shorter wavelength emissions are mostly produced within 10 $r_S$, where electrons are 
relativistic, as indicated by the profiles of $\nu_E$ and $T_e$, the emission and absorption are dominated by the 
extraordinary emission component and the Faraday conversion dominates the Faraday rotation in the millimeter and 
sub-millimeter bands (Melrose 1997). Then the emission, absorption, and dispersion vectors defined by Landi Degl'Innocenti 
and Landolfi (2004) are in the same direction. With the slim disk approximation, the plasma is uniform along a light ray and 
the dispersion effects can be neglected. So the formulae given by Melia et al. (2001) remain valid. Considering the vertical 
structure of the disk and the inhomogeneities caused by turbulence may reduce the level of linear polarization and induce 
circular polarization as implied by equations 5.74a and 5.74b of Landi Degl'Innocenti and Landolfi (2004). 

A recent study of the magneto-rotational-instability indicates that $\beta_\nu\simeq 0.7$ when the turbulence saturates 
(Pessah et al. 2006). In principle, $\beta_p$ can also be determined by realistic MHD simulations. Then 
the observed spectrum and polarization can give tight constraints on the disk inclination angle $i$ and $C_1$.
The right panel of Figure \ref{f1.eps} compares the model predicted spectrum and linear polarization with observations. 
The thick dashed line in the top panel gives the spectrum of the fiducial model with $i=50^\circ$, which fits the 
millimeter to NIR spectrum. The spectral index below $100$ GHz is $\sim 1.7$, which is determined by the structure of the 
disk. Emission from different radii peaks at different frequencies, above which the source becomes optically thin, and lower 
frequency emission is produced at relatively larger radii. The dotted line fit the low frequency spectrum with $F_\nu = 
(\nu/10^{9.5}{\rm Hz})^{0.3}\exp[-(\nu/10^{10.5}{\rm Hz})^{1/3}]$ Jy. As mentioned in the introduction, this component is 
likely associated with an outflow and its linear polarization is negligible (Bower et al. 2002). The thick solid line is the 
sum of the two, which fits the broadband spectrum. 

Although most of the emission from the disk is produced in the optically 
thin regime, the linear polarization at lower frequencies is suppressed due to the dominance of the unpolarized 
component. The polarization is significant only in the millimeter and shorter wavelength range as shown in the middle panel, 
where the thick line corresponds to the fiducial model, and the two thin lines have $i=40^\circ$ and $60^\circ$. 
The polarization fraction in the sub-millimeter range increases dramatically with the increase of $i$. Our 
fitting to the observations gives an $i$ of $\sim 50^\circ\pm10^\circ$. More accurate polarization observations 
above $300$ GHz may give a better measurement of this angle.  


The thin dashed lines in the top panel give the spectra for the two profiles with $C_1=0.36$ and $0.40$. The emission 
spectrum, especially in the NIR band, is very sensitive to the electron heating rate. A good measurement of Sagittarius A*'s 
quiescent flux in the NIR band can lead to an accurate determination of $C_1$ (Ghez et al. 2005b; Cl\'{e}net et al. 2004). 
The model predicts an NIR spectral index of $\sim -3.5$. Any evidence of a harder quiescent NIR spectrum would suggest a 
non-thermal emission component.


Observations show that the position angle of the intrinsic polarization is $\sim 170^\circ$ (Macquart et al. 2006; Marrone et 
al. 2007). Our model predicts that the emission is polarized along the axis of the disk below $\sim 1200$ GHz. The axis of 
the disk therefore has a position angle of $\sim 170^\circ$. Future VLBI imaging may be able to test this prediction (Shen et 
al. 2005; Bromley et al. 2001; Huang et al. 2007). For small values of $i$, the model predicts that the position angle of the 
polarization flips by $90^\circ$ in the sub-millimeter range due to the dominance of the high frequency emission by the blue 
shifted side. We note that in the NIR band, where the emission is also dominated by the blue-shifted side, the corresponding 
position angle of the polarization is $\sim 80^\circ$, which is consistent with recent observations (Eckart et al. 2006b; 
Meyer et al. 2006). The bottom panel shows the fitting to the position angle of the observed linear polarization. Here an 
external Faraday rotation has been introduced following Macquart et al. (2006) and Marrone et al. (2007). Such an external 
Faraday rotation may originate from the outflow as suggested by Beckert and Falcke (2002).

\section{Discussion and Conclusions}
\label{dis}

We have shown that the disk model can account for the millimeter spectrum and polarization, the $\sim 90^\circ$ difference in 
the position angle of the polarized sub-millimeter and NIR emissions. Our best-fit model predicts a $\sim 8\%$ 
polarization at 150 GHz, a very soft quiescent state NIR spectrum, and a disk inclination angle of $50^\circ$ with the axis 
aligning at $\sim 170^\circ$ position angle. Future polarimetric spectroscopy and imaging will be able to test the model 
and/or better constrain the parameters.

Since $\beta_p$ is currently not well determined by simulations, the fitting to the observed data is not unique. For 
$\beta_\nu=0.7$, $\beta_p=0.1$, $C_1 = 1.08$, $\dot{M} = 2\times 10^{17}$g s$^{-1}$, and $i=50^\circ$, we obtain essentially 
the same emission spectrum and polarization from the disk as the fiducial model. A better constraint of $\beta_p$ by 
simulations is therefore critical to pin down the rest of the model parameters with simultaneous spectrum and polarization 
observations. The best-fit orientation of the disk is not sensitive to $\beta_p$. However, since most of the emission is 
produced near the last stable orbit $3r_S$, the stress at this radius and the emission from the plunging region may affect 
the best-fit value of $i$. A full general relativistic MHD simulation will not only address these issues and remove the 
parameters $\beta_\nu$ and $\beta_p$, but also treat the vertical structure of the disk and the effects of winds or outflows 
properly, which may explain the observed external Faraday rotation and the centimeter emission component (Liu \& Melia 2001; 
Yuan et al. 2002; Sharma et al. 2007b).
However, to apply MHD simulations to Sagittarius A* directly, one needs an advanced treatment of the electron heating 
by turbulence, takes into account the general relativistic light binding effects, and does the transfer of polarized emission 
in a turbulent flow properly. The plasma dispersion effects may induce circular polarization and introduce comparable 
modification to the spectrum and linear polarization as the general relativistic effects (Melrose 1997; Broderick \& 
Blandford 2004). These investigations may eventually lead to a measurement of the black hole spin.

Compared with the previous models of a small hot accretion torus (Melia et al. 2001; Bromley et al. 2001), we 
make several significantly improvements. A more realistic treatment of electron heating by turbulence replaces 
the assumption of one temperature flow in the original model. This also makes the results almost independent of 
the outer boundary radius and temperature(s). With the pseudo-Newtonian potential, the inner radius is not a 
free parameter anymore and fixed at the last stable orbit. The zero stress inner boundary condition is replaced by the zero 
angular momentum flux condition. Although other models may also explain the observations studied in this paper (e.g. Yuan 
et al. 2002), the model presented here has less parameters and assumptions on the dynamics of the accretion flow and/or 
outflow and therefore likely reveals the dominant physical processes responsible for the observed emission.

\acknowledgments

This work was funded in part under the auspices of the U.S.\ Dept.\ of Energy, and supported 
by its contract W-7405-ENG-36 to Los Alamos National Laboratory.

\clearpage


\end{document}